Topical Review

# On the potential of hard ferrite ceramics for permanent magnet technology—a review on sintering strategies

Cecilia Granados-Miralles[1]  and Petra Jenuš[2]

[1] Electroceramic Department, Instituto de Cerámica y Vidrio, CSIC, Madrid, Spain
[2] Department for Nanostructured Materials, Jožef Stefan Institute, Ljubljana, Slovenia

E-mail: c.granados.miralles@icv.csic.es

## Abstract
A plethora of modern technologies rely on permanent magnets for their operation, including many related to the transition towards a sustainable future, such as wind turbines or electric vehicles. Despite the overwhelming superiority of magnets based on rare-earth elements in terms of the magnetic performance, the harmful environmental impact of the mining of these raw materials, their uneven distribution on Earth and various political conflicts among countries leave no option but seeking for rare-earth-free alternatives. The family of the hexagonal ferrites or hexaferrites, and in particular the barium and strontium M-type ferrites ($BaFe_{12}O_{19}$ and $SrFe_{12}O_{19}$), are strong candidates for a partial rare-earth magnets substitution, and they are indeed successfully implemented in multiple applications. The manufacturing of hexaferrites into dense pieces (i.e. magnets) meeting the requirements of the specific application (e.g. magnetic and mechanical properties, shape) is not always straightforward, which has in many cases hampered the actual substitution at the industrial level. Here, past and on-going research on hexaferrites sintering is reviewed with a historical perspective, focusing on the challenges encountered and the solutions explored, and correlating the sintering approaches with the magnetic performance of the resulting ceramic magnet.

(Some figures may appear in colour only in the online journal)

## 1. Introduction

Rare-earth elements (REEs), often referred to as simply rare-earths, are the basis of both well-implemented and emerging modern technologies [1]. Although REEs are not as scarce as their name suggests, they are very scattered, and their mining is costly and contaminant [2]. Furthermore, deposits which are worth mining (with sufficient concentrations) are not evenly distributed throughout the globe, with China being the world-leading REE-producer and processor by far, accounting for more than 70% of the total produced [3]. China has already used its dominant position in the past, when they limited the exports in 2009 and 2010, and raised the prices in 2011 [4]. After this event, great powers such as Japan, the EU or the US have made great efforts to reduce REE-consumption and minimize the dependence on China [5–7]. Considering the ongoing US–China trade conflict and the current global COVID-19 crisis [8], the situation threatens to become critical again. Some have anticipated that the Chinese could take revenge on the US by restricting the REE-supply, especially when even China itself is expecting to be affected by shortage [9]. Permanent magnets (PMs) make a textbook example of technology stricken with the REE-problem [10].

Owing to their ability to interconvert between motion and electricity, devices which operation relies on PMs are nearly innumerable. Some examples are: microphones and speakers of everyday electronic devices (mobile phones and computers), hard disk drives and other information storage devices, motors and generators of electric vehicles, machinery to harvest renewable energy sources (e.g. wind or underwater turbines) or household appliance, small motors and sensors in cars (windscreen wipers and power windows motors, steering wheel position sensors) [11]. For many of these applications, reducing the weight is either beneficial or adds value to the product (miniaturization is highly appreciated in our high-tech times). The better the magnetic performance of a magnet, the less amount of magnetic material is required to develop a specific magnetic work. Thus, mass/volume of the devices may be reduced by employing better magnets [12].

At present, Nd–Fe–B magnets are the best known PMs, by virtue of the REE Nd [13]. This intermetallic phase was simultaneously discovered in 1984 by Croat (General Motors Research Laboratories, US) [14] and M Sagawa (Sumitomo Special Metals Company, Japan) [15], and it meant a breakthrough in the field of PMs. Although magnets based on the Sm–Co alloy were already well-developed at the time [16], they were produced in very small quantities (and still are), only reserved to very specific applications, as a consequence of the elevated price of the material [17]. In these REE-based intermetallic compounds, the transition metal (Fe, Co) ensures a high magnetization while the REE (Nd, Sm) is responsible for a very high magnetocrystalline anisotropy, which in turn produces a large coercivity, $H_c$ [18]. As a result, REE-based magnets have an outstanding magnetic performance, with very high energy products ($BH_{max}$), typically above 200 kJ m$^{-1}$ [3, 19] Unfortunately, as mentioned earlier, REEs are subject to irregular market fluctuations and intermittent shortage, which leads to a high dependency on external factors that magnet manufactures and end-users are very interested in avoiding.

Two main options are currently being explored in order to overcome this problem: (a) REE-recycling and (b) REE-substitution. The most straightforward approach has been raising awareness and implementing strategies towards the recycling and reuse of REEs [10, 13, 20]. However, this can only function as a short-term solution. Positively preventing the criticality of raw materials from hindering further technological advances unarguably requires giving consideration to the full (or at least partial) substitution of REE-based magnets. This is undoubtedly a more laborious process but shall be worthy if outlined as a long-term solution.

Although REE-magnets are indispensable in some high-performance devices, there are many others not requiring so much magnetic power but where Nd–Fe–B magnets are still employed, since the just-below alternatives are considerably worse in performance and cannot meet the requirements [21]. In many of these scenarios, subtle changes on the engineering of the device, such as reinventing the magnets assembly and/or rethinking the device design, allow for a smaller magnetic power demand, which automatically makes substitution possible [22–24]. There are also quite some cases for which a moderate betterment of the magnetic properties would allow replacing REEs by lower-grade magnets without exceeding the mass limitation. Thus, great efforts are devoted to shrinking the gap between REE-containing and REE-free magnets, either by finding new REE-free alternatives or by optimizing the performance of known magnetic phases [25].

Hexagonal ferrites have long been considered as plausible 'gap magnets' for REE-substitution, as a consequence of their relatively good magnetic properties coming along with a remarkably high stability and high Curie temperature, all this at a very low price compared to REE magnets [26]. However, manufacturing dense hexaferrite pieces (i.e. hexaferrite magnets) with high enough densities while maintaining the good magnetic properties of the powders has proven not straightforward, which in many cases has hampered a marketable REE substitution by ferrites. The technical difficulties encountered during the densification of hard ferrites over the years and the numerous studies addressing this topic stand as the motivation for the present review.

## 2. Hexagonal ferrites

### 2.1. General characteristics and properties. Why hexagonal ferrites?

Hexagonal ferrites, also known as hexaferrites or simply hard ferrites, were first announced as PM materials in 1952 by van Oosterhout and co-workers at the Philips Research Laboratories (Eindhoven, The Netherlands) [27, 28], based on the former work on magnetic oxides performed by Snoek, also at Philips [29]. This family of compounds are ternary or quaternary iron oxides that crystallize in a hexagonal lattice with considerably long $c$-dimension (23.03–84.11 Å) [30]. Among all hexaferrites, the so-called M-type are particularly interesting for PMs applications, owing to the large magnetocrystalline anisotropy along the crystallographic $c$-axis, derived from their atomic structure. The strong uniaxial anisotropy causes the M-type ferrites to have fairly large theoretical maximum $H_c$, making them very robust against demagnetization (magnetically hard) and therefore very useful as PMs.

The M-type hexaferrites have general formula $M\text{Fe}_{12}\text{O}_{19}$, $M = \text{Ba}^{2+}$ or $\text{Sr}^{2+}$, and are typically abbreviated as BaM and SrM. They crystallize in a hexagonal magnetoplumbite structure ($P6_3/mmc$) with two formula units per unit cell. The crystal lattice has very anisotropic dimensions ($a = 5.892$ Å, $c = 23.18$ Å for BaM and $a = 5.884$ Å, $c = 23.05$ Å for SrM) and the crystallographic (theoretical) densities are 5.30 and 5.10 g cm$^{-3}$, respectively. [31, 32] The lattice can be divided into two alternating structural blocks stacked along the $c$-direction (cubic S and hexagonal R block, respectively)

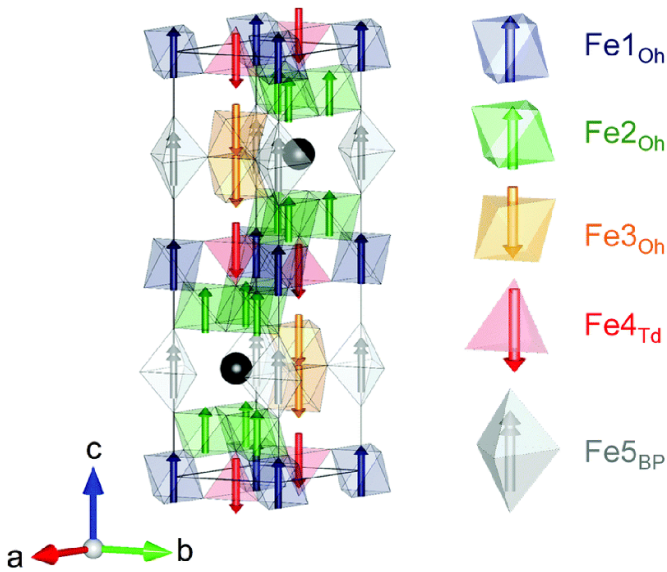

**Figure 1.** Representation of the crystal and magnetic structure of the M-type hexaferrites, i.e. $BaFe_{12}O_{19}$ and $SrFe_{12}O_{19}$. The black spheres represent the alkaline earth metals (Ba, Sr) while the colored polyhedra represent the five different Fe crystallograpic sites. The arrows represent the magnetic spins of the Fe atoms. Oxygen atoms are found at all the vertices of the Fe polyhedra. Reproduced from [35] with permission from the Royal Society of Chemistry.

[33]. The basic unit cell is composed of two R- and two S-blocks, RSR*S*, where * denotes a 180° rotation around the *c*-axis, and contains two formula units (64 atoms in total) [33]. The $Ba^{2+}$ (or $Sr^{2+}$) and $O^{2-}$ are arranged in a close-packed manner, while the smaller $Fe^{3+}$ ions are located at interstitial positions [30, 33, 34]. The crystal (and magnetic) structure of M-type hexaferrites is shown in figure 1.

The intrinsic magnetic parameters of BaM and SrM are listed in table 1 and they will be discussed below. The net magnetic moment of M-type hexaferrites arises from the antiferromagnetic alignment of the individual $Fe^{3+}$ magnetic moments (see arrows in figure 1), resulting in a theoretical (at 0 K) magnetic moment of 20 $\mu_B$ per molecule of $BaFe_{12}O_{19}$ and 20.6 $\mu_B$ for SrM [30, 36]. This translates into rather high maximum (saturation) values of mass-magnetization, $M_s$, and magnetic flux density or magnetic induction (per unit volume), $B_s$. Their Curie temperature, $T_C$, is outstanding compared to most magnets, including REE-based. Thus, the $T_C$ of BaM or SrM is more than 150 °C higher than that of $Nd_2Fe_{14}B$.

Both compounds have relatively high anisotropy constants, $K_1$, which gives them a rather large magnetocrystalline anisotropy, $H_A$, along the *c*-axis (see table 1) [37–39]. In a single-crystal form, the maximum $H_c$ values reported are 594 kA m$^{-1}$ for BaM and around 533–597 kA m$^{-1}$ for SrM [28, 33, 37, 38]. However, when in a polycrystalline form, the coercivity of M-hexaferrites largely depends on the particle size and it is usually much smaller than the theoretical value [28, 30, 33]. As a side note, the coercivity of BaM and SrM display an interesting behaviour with temperature: for both compounds, $H_c$ increases with temperature up to a certain value (for polycrystalline BaM, the peak is reached at approximatelly 250 °C with $H_c$ of 380 kA m$^{-1}$) [28, 33]. Beyond this certain value, $H_c$ decreases with increasing temperature, which is the behaviour usually found in ferrimagnetic materials.

Despite the overwhelming magnetic superiority of Nd-based magnets, hexaferrites have been widely used in the PMs industry since their discovery. They are still of great research interest as evidenced by the record number of publications on the topic [30], and even today, hexaferrites continue to be the most produced magnetic material [40]. The hegemony of hard ferrites compared to REE-magnets, despite the superior performance of the latter, rely on their decent performance at a very low price. For instance, in 2013 hexaferrites accounted for 85 wt% of the PMs sales, although they only represented 50% of the billing [18]. The cost of a raw material is generally related to the abundance of the elements in the Earth's crust, and the elemental constituents of hexagonal ferrites are widely available. Fe is the fourth most abundant element in Earth's crust, and the indisputable first among the magnetic elements [39]. In particular, Fe is more than a thousand times more abundant than Nd [41], and a thousand times cheaper [40]. Oxygen is the most abundant element while Ba and Sr are in 14th and 15th positions, respectively [41]. Apart from the price, hexaferrites present other advantages compared to REE-based magnets. They have an outstanding chemical stability and resistance to corrosion, and their high $T_C$ is very convenient when utilized in motors or other devices that tend to acquire temperature during operation.

Like for any other functional material, the magnetic properties of M-type hexagonal ferrites do not only depend on their structure and chemical composition [38], or the temperature of operation [37]. They are also highly influenced by the method used for the synthesis of the magnetic powders [30, 42–50], as well as by the strategies followed to consolidate the powders [30, 33, 51–55], and very importantly, by the magnetic alignment or orientation of the densified piece [28, 33, 56, 57]. The last one is particularly interesting here, as the particles of the M-type hexaferrites have a tendency to acquire a plate-like shape, with the platelet plane perpendicular to the *c*-axis (see figure 2) [58, 59]. This shape favors a crystallographic alignment during compaction under a uniaxial pressure, during which the plate-like particles will have a tendency to spontaneously lay on their larger surface side, just as a deck of cards would. This shape-induced orientation implicitly entails a magnetic alignment, as the easy magnetization axis lies parallel to the *c*-axis (perpendicular to the platelet plane) [52, 58], vanishing the need for an external magnetic field applied prior or during compaction, which considerably simplifies manufacturing the magnetic piece. Preparation methods and magnetic properties of M-type ferrites in the powder shape are extensively reviewed in another article of the present issue [26], while aspects related to consolidation and magnetic orientation of the powders are addressed here.

### 2.2. General aspects on consolidation and typical problems encountered

In order to be an integral part of any device, magnetic powders need to be conformed into dense pieces (i.e. magnets) with specific shapes and dimensions. Like with any other material, a good density is required to ensure good mechanical performance of the component. But this is even more crucial in the case of magnetic materials, given that the magnetic figure of merits ($BH_{max}$) scales with the density. Simplistically speaking, the densification of powders is generally carried out by either simultaneous or consecutive application of pressure and temperature. This single- or multiple-step consolidation process is known as sintering, and in particular, when applied to ceramic materials such as ferrites, it may be referred to as ceramic sintering [60].

**Table 1.** Basic intrinsic magnetic properties of BaFe$_{12}$O$_{19}$ and SrFe$_{12}$O$_{19}$, along with those of Nd$_2$Fe$_{14}$B for comparison. All temperature-sensitive parameters are room temperature values [33, 37–39].

|  | $M_s$ (Am$^2$ kg$^{-1}$) | $B_s$ (T) | $T_C$ (°C) | $K_1$ (MJ m$^{-3}$) | $H_A$ (MA m$^{-1}$) |
|---|---|---|---|---|---|
| BaFe$_{12}$O$_{19}$ | 72 | 0.48 | 467 | 0.32–0.33 | 1.34–1.35 |
| SrFe$_{12}$O$_{19}$ | 74.5 | 0.48 | 473 | 0.35–0.36 | 1.47–1.59 |
| Nd$_2$Fe$_{14}$B | 165 | 1.61 | 315 | 4.9 | 6.13 |

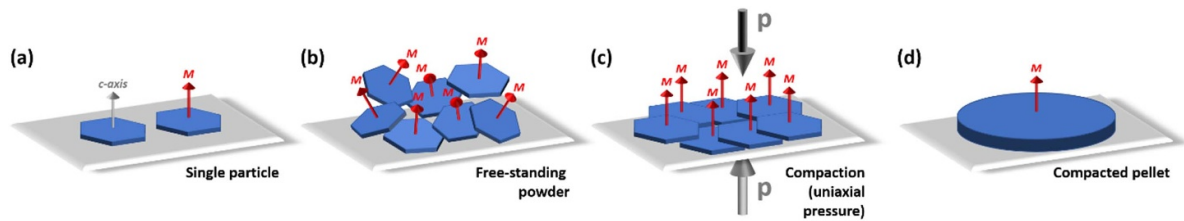

**Figure 2.** Illustration of the shape-induced orientation occurring during compaction of M-type platelets under uniaxial pressure. The magnetic easy axis (i.e. the net magnetization direction) lies parallel to the crystallographic c-axis.

In the consolidation of functional materials, which are optimized in the powders shape, it is crucial that the material properties remain unaltered after sintering. In fact, great efforts are devoted to either adapt the conventional routes or develop new sintering strategies which are optimized for each specific material, so that the damage is minimized [60]. As a matter of fact, sintering of hexaferrites often ends up in a loss of magnetic attributes [61]. A frequent problem when sintering hexaferrites is the appearance of secondary phases, many times the highly stable iron oxide $\alpha$-Fe$_2$O$_3$ (hematite), which antiferromagnetic character diminishes the saturation magnetization. Luckily, when the starting powders are stoichiometrically pure and the thermal treatment is complete, this is generally not an issue, and the $M_s$ values attained experimentally are usually around 70 Am$^2$ kg$^{-1}$, indeed close to the maximum theoretical values [30].

Substantially more difficult to avoid is the damage to the $H_c$ resulting from the elevated temperatures required to attain good densities in the sintered piece. Hexaferrites are generally sintered between 1100 °C and 1350 °C [30], and these high temperatures trigger an exaggerated grain growth which translates into a dramatic loss in coercivity [62]. Thus, while for hexaferrite powders, $H_c$ values close to the single-crystal maxima have been reported (487–525 kA m$^{-1}$) [49, 63–65], the standard ceramic methods usually yield much lower values of around (<300 kA m$^{-1}$), as a consequence of the grain growth during the process [30, 61]. Thus, the $BH_{max}$ values obtained after consolidation of BaM and SrM powders is generally far below the theoretical maximum of 45 kJ m$^{-3}$ predicted for these compounds [17]. Different strategies are pursued to retain $H_c$ after sintering, such as fine-tuning the thermal cycles in order to maximize the density while minimizing the temperature [66–68], or including different additives to hinder grain growth [69–71].

Another common strategy to minimize the $H_c$ loss is mixing the magnetic powders with a polymer matrix and manufacturing the mixture to obtain a dense piece or bonded magnet [73, 74]. The presence of the polymer reduces $M_s$ compared to a pure ceramic magnet, and the maximum operating temperature in the applications may be limited by the glass transition or the viscosity temperature of the polymer (mechanical softening), but in return, bonded magnets can be easily made into any desired shape, widening the spectrum of applications of ferrites (sintered pieces are generally limited to simple shapes). Regarding the mechanical properties, bonded ferrites are generally better in terms of flexural strength (more ductile), while ceramics have much higher hardness. The details on bonded ferrites are out of the scope of this review and are gathered elsewhere [75–81]. Here, different ceramic sintering techniques applied to hexaferrites are reviewed, analyzing the problems encountered, the proposed solutions and the main advances introduced during the last decades. The quality of the obtained ceramic magnets is evaluated in terms of the density and the magnetic performance.

## 3. Sintered hexaferrites

Sintering is one of the key procedures in manufacturing and processing of ceramics. It is the step responsible for the transformation of particulates to net-shaped bodies with the desired properties. The material densification during sintering occurs as a consequence of matter transport to or around pores, triggered by appropriate conditions of temperature, pressure and atmosphere [82–84]. Sintering is usually accompanied by changes in the microstructure, such as reduction in the content and size of pores and grain growth. Although sintering itself is an ancient process that has been carried out for thousands of years, the scientific investigation on the phenomena taking place during sintering has mostly been in progress from the 1950s [84]. Besides continues optimization of the conventional ceramic methods, a plethora of new sintering methods have been explored in the last decades, aiming at facing new challenges such as microstructure preservation after sintering [60]. This is no exception for ferrites.

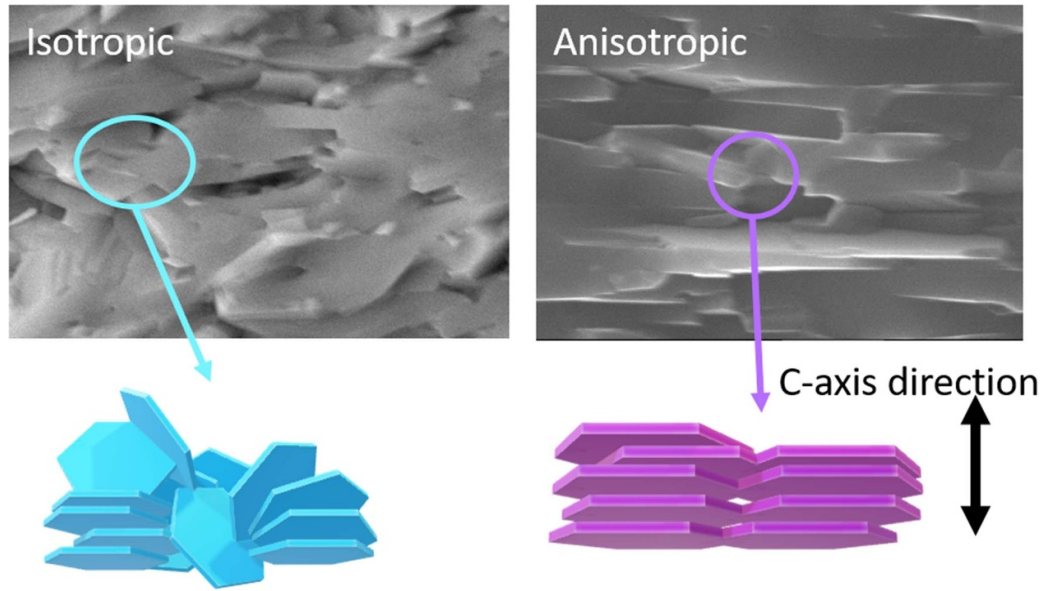

**Figure 3.** SEM micrographs and schematic representation of iso- and anisotropic sintered hexaferrite magnets. Reprinted by permission from Springer Nature Customer Service Centre GmbH: Springer Nature, Journal of Materials Science: Materials in Electronics [72], Copyright 2013.

The history of sintered ferrites started almost 100 years ago with Takei and Kato from Tokyo Institute of Technology (Japan) and their patent on production of soft magnetic zinc-ferrite cores [85], which led to the TDK Corporation and with that, to the commercialization of ferrites. Shortly after, European researchers from Philips Research Laboratory (The Netherlands) also started investigating ferrites. They initially focused on soft ferrites but they later went into hard ferrites too, and in 1952 Philips came out to announce the first commercial hard ceramic magnet, which was composed of hexagonal Ba-ferrite and to which they referred as Ferroxdure [27, 28]. Since then, there has been, and still is, a lot of research focused on the consolidation of hexaferrites, in a quest to produce ferrite PMs with improved magnetic properties [86].

For PM materials, the important properties are the coercivity, $H_c$, and the remanent induction (or simply remanence), $B_r$, targeting the highest possible maximum energy product, $BH_{max}$. The coercivity strongly depends on the microstructure of the sintered material, and especially on the grain size. If the grains contain a single magnetic domain, the demagnetization through domain wall motion is not possible (as there are no domain walls in a single-domain body) and therefore, the demagnetization of the piece can only occur by domain rotation, which is fairly difficult in a dense material, which in turn results in a high $H_c$ [18, 19, 30, 33]. In most ferrites, the particles/grains are in a single domain state when their size does not exceed one micron [33]. On the other hand, the key parameters determining the remanence are the chemical composition, the density and the degree of grain orientation within the sintered magnet, i.e. the purest, the denser and the more magnetically-oriented, the higher the $B_r$ value of the magnet.

At present, BaM and SrM magnets are commercially available in both their non-oriented (isotropic) and oriented (anisotropic) forms, and the difference in $B_r$ (and therefore $BH_{max}$) between them is remarkable. While commercial isotropic (randomly oriented) ferrite magnets display energy products up to about 10 kJ m$^{-1}$ [3, 87, 88] anisotropic (magnetically oriented) magnets reach up to 33.0–41.8 kJ m$^{-3}$ [87–90]. Higher values (up to 44 kJ m$^{-3}$) are found for La- and Co-substituted ferrites [91, 92]. Scanning electron microscopy (SEM) images of both types of sintered hexaferrite magnets are shown figure 3, along with the corresponding schematic representation of both cases.

Although sintered ferrite magnets available on the market are consolidated by the conventional method, optimization of the traditional ceramic routes is still an active research topic. Besides, the application of novel sintering techniques, such as microwave or spark plasma sintering (SPS), to ferrites is also a rich and interesting field. In the following sections, the most relevant advances on M-type ferrites sintering are highlighted.

### 3.1. Conventional sintering

The original *Ferroxdure* phase reported by Went *et al* in 1952 was an isotropic Ba-ferrite with a high coercive force ($\approx$240 kA m$^{-1}$) and a relatively low remanent induction (0.2–0.21 T), which resulted in a maximum energy product of 6.8 kJ m$^{-3}$ [27, 28]. Although these first two papers about Ferroxdure presented its magnetic properties in detail, they did not describe the preparation procedure and the conditions used for sintering. In the following paper on Ferroxdure, authored by Stuijts *et al* the development of anisotropic Ba-hexaferrite magnets with energy products up to 28 kJ m$^{-3}$ was announced [93]. To prepare an anisotropic ferrite magnet, they used randomly oriented non-aggregated Ba-ferrite powder, which was pressed-formed as an aqueous sludge under an applied external magnetic field. The excess water was then removed by drying followed by the sintering of the pre-forms. This process developed in Philips Research Laboratories in 1950s is, with some modifications, still in use for the preparation of commercial sintered ferrite magnets.

Stuijts *et al* tested sintering temperatures ranging from 1250 °C to 1340 °C, and they found out that increasing the sintering temperature yields a higher degree of orientation [93]. Thus, an almost fully anisotropic magnet was obtained at the highest sintering temperature (1340 °C). On the contrary, the coercivity was the lowest (<20 kA m$^{-1}$) at this temperature. They also found out that the grain growth occurring during sintering enhances the magnetic anisotropy, but at the same time, it limits the $BH_{max}$ due to the decrease in coercivity. In the same paper, they calculated the maximum value of $BH_{max}$ for a Ba-ferrite single crystal, assuming a 100% degree of orientation, resulting in a remanence of 0.42 T and consequently, a $BH_{max}$ of 35 kJ m$^{-3}$.

The coercivity of ferrite magnets largely depends on their grain size. Namely, the highest coercivities are reached when the grain size is in the narrow region around the limit of the single domain [27, 38, 94]. The final stages of sintering are generally accompanied by an (extensive) grain growth, which is even more pronounced when powders with small particle sizes are densified [84, 95, 96]. As an example, the study by El Shater *et al* brings out the pronounced damage to the coercivity caused by the elevated temperatures and prolonged times when sintering nano-sized BaM [61]. Thus, pellets made out of the same batch of powders sintered at 1000 °C and 1300 °C presented coercivity values of 271 and 56 kA m$^{-1}$, respectively.

The densification of final stages is not at all unwanted, as the remanence of a magnet increases linearly with the density, so the optimal sintering method should ensure a trade-off between a controlled grain growth and a good densification. In a quest for the suppression of grain growth during final stages of sintering without hampering densification, various sintering additives have been used, SiO$_2$ being one of the most widely explored [57, 71, 94, 97–100]. In 1985, Kools concluded that the highest $H_c$ were obtained for SiO$_2$ concentrations within the range 0.36–1.44 wt% and he put forward a mechanism for the grain growth impediment [97, 98]. In 1991, Beseničar and Drofenik studied the sintering of fine SrM particles with a small addition (0.5 wt%) of SiO$_2$ [99]. They found out that SiO$_2$ does not only suppress grain growth, but it also promotes rearrangement of Sr-ferrite particles during sintering leading to highly anisotropic magnets with elevated density (97% of the theoretical value). These values were achieved at 1260 °C with a holding time of only 3 min at the final temperature, and yield a remanence of ≈0.39 T and a coercivity of ≈340 kA m$^{-1}$ (graphically estimated from figure 5 in [99]). In 2013, Kobayashi *et al* concluded that to improve the coercivity, the amount of added SiO$_2$ is preferably 1–1.8 wt%, while additions higher than 1.8 wt% undesirably lower the coercivity [100].

While small additions of SiO$_2$ suppress ferrite grain growth, the addition of CaO promotes densification and consequently increases the remanence [57, 94, 100]. The downside of the addition of CaO is the extensive grain growth, which negatively affects the coercivity (and also $BH_{max}$). However, a combination of the two additives in the right amounts has shown good results [101, 102]. In 1999, Lee *et al* obtained an enhanced coercivity of 281 kA m$^{-1}$ by sintering SrM mixed with 0.6 wt% SiO$_2$ and 0.7 wt% CaO at 1200 °C for 4 h [101]. However, the orientation was not so high, which yield only moderate values of remanence (0.36 T) and $BH_{max}$ (29.4 kJ m$^{-1}$ [3]). In 2005, Töpfer *et al* investigated the effect of simultaneous addition of SiO$_2$ and CaO on the microstructure and magnetic properties of sintered Sr-ferrite [102]. They found out that a Sr-ferrite powder mixed with 0.5 wt% of SiO$_2$ and CaO in a 1:1 ratio, and sintered at 1280 °C for a short period of time, resulted in a dense Sr-ferrite magnet (98%) with refined microstructure and good magnetic properties ($B_r$ = 0.42 T, $H_c$ = 282 kA m$^{-1}$, and $BH_{max}$ = 32.6 kJ m$^{-3}$). In 2018, Huang *et al* tested the simultaneous addition of CaCO$_3$ and SiO$_2$ together with Co$_3$O$_4$ [70]. By conventionally sintering a mixture of SrM powders with 1.1 wt% CaCO$_3$, 0.4 wt% SiO$_2$, 0.3 wt% Co$_3$O$_4$, and 0.5 wt% of a dispersant, they managed to manufacture a 99% dense piece with remarkably good magnetic properties ($B_r$ = 0.44 T, $H_c$ = 264 kA m$^{-1}$, and $BH_{max}$ = 38.7 kJ m$^{-3}$).

Besides the use of sintering additives, other approaches have been explored to overcome the grain growth problem. Following a two-step sintering method proposed by Chen and Wang in 2000 which offered good results on nanosized Y$_2$O$_3$ [95], Du *et al* recently presented a two-step sintering profile adapted to Sr-ferrite [66]. In this work, the green body is first heated to a higher temperature for a short time (1200 °C for 10 min) and then immediately cooled to an intermediate temperature, at which it remains to continue sintering for a prolonged time (1000 °C for 2 h). Such a procedure resulted in a fully dense Sr-ferrite magnet with a remanence of 0.44 T, a coercivity of 328 kA m$^{-1}$, and a $BH_{max}$ of 37.6 kJ m$^{-3}$.

### 3.2. Microwave sintering

While in the majority of cases microwaves (MW) are used as a synthesis aid [30, 46, 47, 57, 103, 104], there are also some reports on MW sintering of hexaferrites [72,105,106]. For microwave sintering, the powders are generally cold-pressed into relatively dense pieces (green bodies) which are then sintered by electromagnetic radiation in the GHz region MW oven [60]. The preliminary tests carried out by Binner *et al* in 1999, suggested that it was possible to MW-sinter nano-sized ferrite powders without inducing much grain growth as long as the starting powders were not agglomerated [105]. Unfortunately, most of the dense pieces they produced contained cracks after the sintering. In 2009, Yang *et al* prepared 97% dense BaM magnets by microwave sintering [106]. They could avoid the presence of α-Fe$_2$O$_3$ as an impurity while reducing the sintering temperature considerably compared to the conventional ceramic methods. However, in this case, they could not avoid grain growth, yielding to a very low $H_c$ (<50 kA m$^{-1}$).

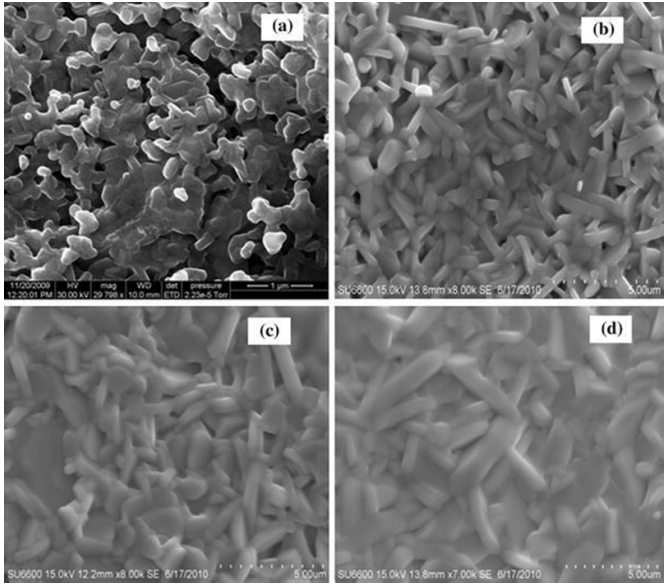

**Figure 4.** SEM images or SrFe$_{12}$O$_{19}$ microwave sintered for 10 min at (a) 1000 °C, (b) 1050 °C, (c) 1100 °C, and (d) 1150 °C. Reprinted from [107], Copyright 2002, with permission from Elsevier.

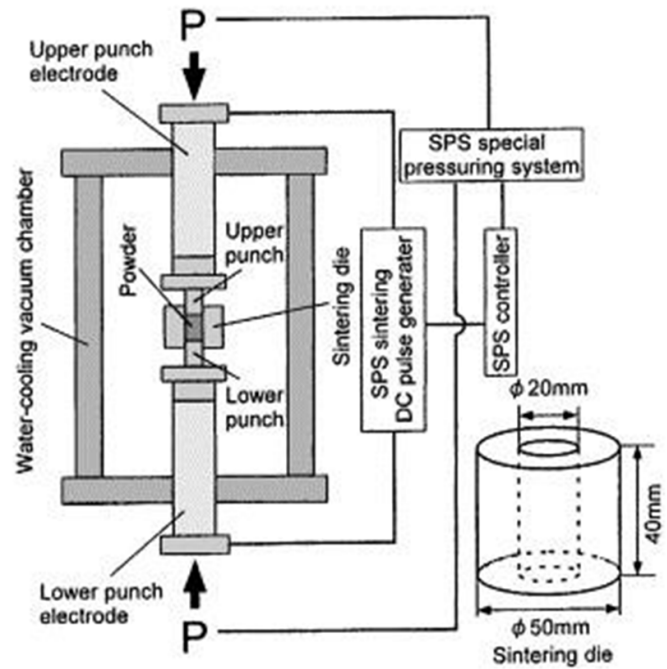

**Figure 5.** Typical spark plasma sintering experimental setup. Reprinted from [51], Copyright 2014, with permission from Elsevier.

Much more promising results were presented by Kanagesan *et al* in 2013 [72]. They used sol-gel synthesized Sr-ferrite, which was uniaxially pressed and sintered by microwave sintering. Figure 4 shows SEM images of the microwaved sintered samples obtained at temperatures between 1000 °C and 1150 °C, with a heating rate of 50 °C min$^{-1}$, and with a holding time at the maximum temperature of only 10 min. At the highest sintering temperature tested (1150 °C), they could make a fairly dense SrM material (95%) with very a competitive $H_c$ of 445 kA m$^{-1}$, although the $B_r$ was not so high in this case.

### 3.3. Spark plasma sintering

With the start of the new millennium, a densification technique called SPS started to gain more and more interest in material science, and also in the field of ferrite magnets. In SPS, a powder sample is introduced in a sintering die, generally made of graphite. The powders are subjected to a uniaxial pressure while an electrical current is run through the sample to heat it up by Joule effect (see figure 5 for a typical SPS setup). This heating modes enables fast heating rates, lower sintering temperatures and shorter sintering times compared to the conventional and microwave sintering [108, 109]. All this features posed a viable option for a densification of materials with no, or with a limited grain growth [109], which would positively affect the coercivity and consequently also the maximum energy product of ferrite magnets. On the other hand, a positive effect on the magnetic properties (especially on the remanence) could also have a direct application of uniaxial pressure during densification by SPS, by causing the preferential orientation of hexagonally shaped grains. So, SPS studies including various M-type hexaferrites and composites based (mainly) on Ba- or Sr-hexaferrite were presented in the last 20 years or so [35, 51–53, 55, 58, 64, 110, 107,111–115], and some of them will be discussed below.

In 2002, Obara *et al* presented results of SPS sintered La, Co-doped SrM ferrite fine particles (1.0 wt% La$_2$O$_3$ and 0.1 wt% Co$_3$O$_4$) [107]. With a final temperature of 1100 °C, a holding time of only 5 min and an applied pressure of 50 MPa, they prepared fully dense samples (density = 5.15 g cm$^{-1}$ [3]) with a remanence of 0.32 T, a coercivity of 325 kA m$^{-1}$, and a $BH_{max}$ of 18.3 kJ m$^{-1}$ [3]. In 2006, Zhao *et al* SPS sintered a textured BaM powder (nanorods) [110]. Although the sintering temperature was rather low (800 °C), the $H_c$ value was not stunning (112 kA m$^{-1}$). In 2011, Mazaleyrat *et al* carried out several SPS tests on pure BaM nanoparticles (< 100 nm) obtaining high $H_c$ values [53]. By limiting the growth of the nanoparticles during SPS, they managed to achieve a coercivity of 390 kA m$^{-1}$ for the dense material. This value is even higher than the obtained by Obara for the La, Co-doped material [107]. However, the magnet was not so dense in this case (88% of the theoretical density), and therefore, the resulting $BH_{max}$, was considerably lower, i.e. 8.8 kJ m$^{-3}$.

In 2014, Ovtar *et al* evidenced the notable smaller sizes obtained after SPS of nanosized BaM powders compared to the same material sintered conventionally (see figure 6) [51]. They studied the partial decomposition taking place during the SPS process and additionally, they noticed that the impurity phases segregate within the sintered material, resulting in the formation of a magnetite or hematite layer at the surface of the sintered pellets while the inside remained Ba-rich. They demonstrated that the protective layer placed between the powders and the graphite die has an influence on the decomposition. After testing different materials (boron nitride, gold, alumina), they obtained their best results using alumina protective discs, yielding an 82% dense pellet with a coercivity of 350 kA m$^{-1}$.

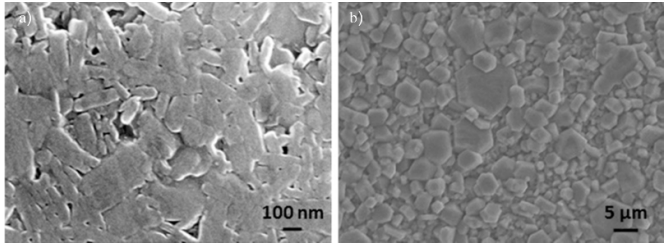

**Figure 6.** SEM images of the inside of (a) an BaM pellet SPS sintered at 900 °C and (b) the same powder conventionally sintered at 1350 °C. Reproduced from [58] with permission of The Royal Society of Chemistry.

In 2015, Stingaciu *et al* tested a ball milling procedure a micro-sized commercial SrM powder aimed at reducing the particle size of the powders prior to SPS [55]. The method was not fully successful as the amorphous phases generated during milling transformed into secondary phases during SPS. Although the sintered densities were well above 90% of theoretical density, the magnetic properties were highly affected by the formation of these secondary phases, which significantly lowered the $BH_{max}$ values of the sintered magnets (values reported were in the range of 3.5–4.6 kJ m$^{-3}$). Later, the $BH_{max}$ could be improved to some extent (up to 9.6 kJ m$^{-3}$) by an additional heat treatment. Thus, it follows that not only pre-treatment of powders used for the consolidation, but also sintering by SPS itself can pose a risk for the decomposition of Ba- or Sr-ferrites.

In 2016, Jenuš *et al* presented a study on the SPS sintering of hard-soft composites based on hydrothermally (HT) synthesized Sr-ferrite [115]. Out of all the samples reported by Jenuš, only the SPS sintering of Sr-ferrite alone is considered here. All sintered SrM samples had densities higher than 90% of the theoretical value, regardless of sintering temperature (700 °C–900 °C) or holding time (1 or 5 min). The best pure SrM pellet from this study was sintered at 900 °C, with a holding time of 5 min and an applied pressure of 92 MPa, and displayed remanence $M_r$ of 65.8 Am$^2$ kg$^{-1}$, a coercivity of 167 kA m$^{-1}$ and a high $BH_{max}$ of 21.9 kJ m$^{-3}$. Although the $H_c$ value is not particularly high in this case, the $BH_{max}$ value is rather large compared to other old studies, and this can be explained by the highly anisotropic morphology of the HT synthesized SrM particles used. The anisotropic shape favors the SrM particles to lie on their basal plane when pressed, this is, with the magnetization pointing out of the pellet plane, which translates in a pronounced magnetic alignment ($M_r/M_s$ ratio[3] = 0.90) and consequently a high $BH_{max}$.

In the same year, Saura-Múzquiz *et al* published another SPS study where the starting SrM powders were HT-synthesized hexagonal platelets with very small sizes (e.g. for some samples the platelet thickness was only slightly over the dimension of one unit cell) [58]. By tuning the SPS routine, they managed to minimize the grain growth during sintering, with which the resulting pellets had a relatively large coercivity of 301 kA m$^{-1}$. High-resolution powder x-ray diffraction data were measured on the SrM powders before and after compaction and Rietveld analysis on those data was carried out to extract quantitative information. Figure 7(a) shows the measured diffraction patterns and Rietveld models of the powders prior to compaction (red) and the densified SPS pellet (blue). Both patterns correspond to the same crystallographic phase, SrM, however, the relative intensities of the peaks are very different, as a consequence of the crystalline orientation. While the powders present a random orientation of the crystallites, a pronounced crystalline alignment (texture) is observed for the pellet. This crystalline alignment was further studied using x-ray pole figures on different Bragg reflections, confirming the preferential orientation along the c-crystallographic direction (see figure 7(b)). This crystalline alignment translates in a pronounced magnetic alignment ($M_r/M_s$ ratio = 0.89), which combined with the high coercivity results in an even higher $BH_{max}$ value of 26 kJ m$^{-3}$. As explained above, and graphically represented in figure 2, the alignment of thin SrM platelets occurs due to the uniaxial pressure applied during the SPS procedure. Furthermore, this spontaneous crystallographic alignment is very convenient from the magnet manufacturing point of view, as the resulting dense pieces exhibit a high magnetic alignment without the need for an externally applied field prior or during compaction, this way avoiding a whole step in the fabrication chain [111]. This characteristic alignment has also been seen to have an effect on the thermal conductivity (much higher along the in-plane than the out-of-plane direction), which opens an opportunity for effective cooling [52].

Further investigations based on both *ex situ* and *in situ*, x-ray and neutron powder diffraction measurements, have allowed to optimize the preparation method (HT synthesis + SPS), with which a $BH_{max}$ of 30 kJ m$^{-3}$ could be achieved by SPS [48, 112, 113, 116]. Applying a magnetic field prior to sintering showed an improvement on the texture ($M_r/M_s$ ratio = 0.95), but this came at the cost of a reduced coercivity (133 kA m$^{-1}$), and the $BH_{max}$ betterment was not significant (29 kJ m$^{-3}$)[114]. Instead, subsequent post-annealing of the SPS pellet (4 h at 850 °C) did allow a great enhancement of the squareness of the hysteresis loop ($M_r/M_s$ ratio = 0.93), reaching a $BH_{max}$ of 36 kJ m$^{-3}$, which would approximately correspond to a 'Y36' in Chinese standards [88, 90]. Similar sintering routes have been tested on SrM powders with different morphology, including commercial powders and powders synthesized following a conventional sol-gel method and a modified sol-gel method [35, 64]. However, the resulting alignment was not as exaggerated as for the thin SrM platelets obtained HT.

---

[3] The common parameter used to estimate the degree of magnetic orientation is the remanence-to-saturation ratio in mass units ($M_r/M_s$), as the maximum possible remanence, $M_r$ (100% oriented magnet) is set by the corresponding saturation value, $M_s$, of the measured M-H hysteresis curve [19].

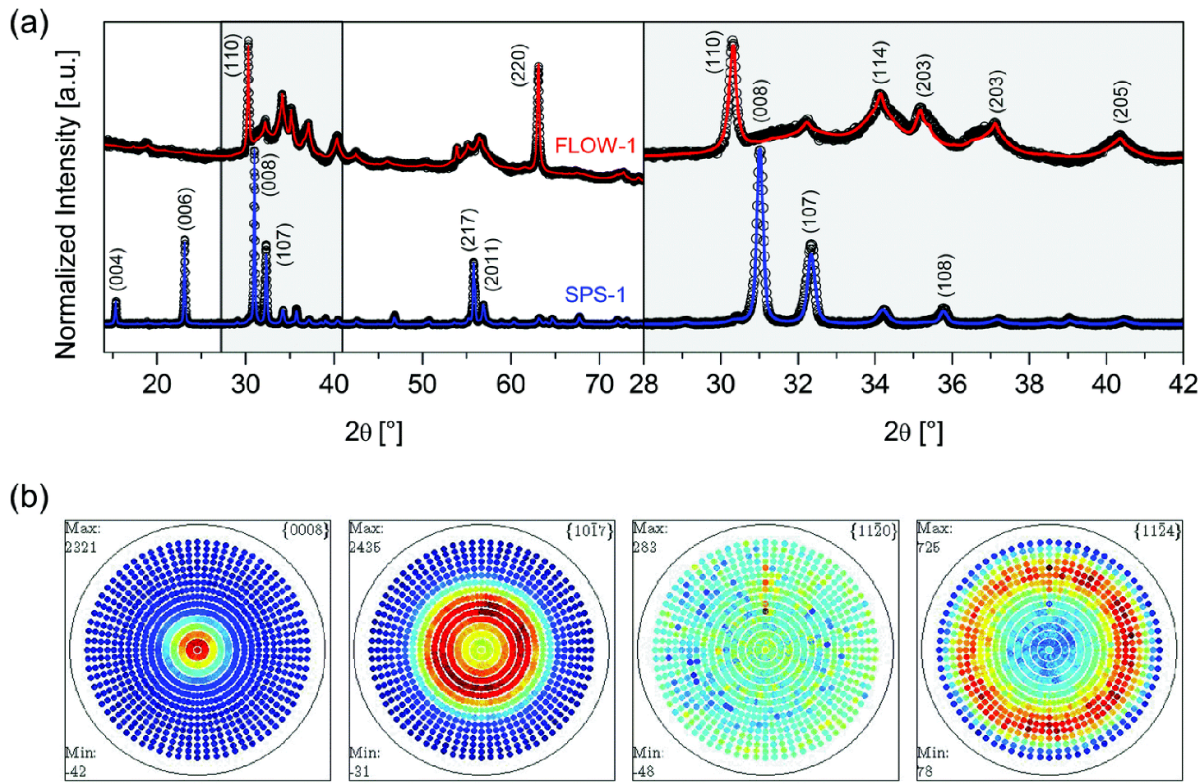

**Figure 7.** (a) Powder x-ray diffraction data along with the corresponding Rietveld models of a hydrothermally-synthesized SrM powder (red line) and a dense SPS pellet fabricated out of those powders (blue line). (b) X-ray pole figure measurements of the SPS pellet. Reproduced from [58] with permission from the Royal Society of Chemistry.

In summary, consolidation of hexaferrites by SPS is following trends valid also for other (ceramic) materials. Namely, the method uses lower sintering temperatures and much shorter sintering times than conventional sintering, and thus the particles' growth is limited. With limited grain growth, the consolidation of nano-sized hexaferrites is effectively resulting in smaller changes in the coercivity when compared to the conventionally sintered magnets, in which grain growth is substantial, as retaining nano-sized particles in the sintered sample usually results in higher coercivities. The uniaxial pressure applied during the process positively affects the density and it also promotes texturing (see figure 2). When dealing with magnetic materials, high density and magnetic alignment lead to an increase in remanence. Thus, a higher degree of texturing is reflected in the higher squareness of the hysteresis loops, which together with high remanence and high coercivity leads to better maximum energy product. Although SPS provides several advantages when it comes to the consolidation of M-type ferrite (nano)powders, there is also a drawback of this method, which needs to be taken into an account. If graphite dies with protecting graphite sheets are used, a chemical reduction of the material at the surfaces may occur. Depending on the reduction degree, the magnetic properties of the sintered piece can deteriorate substantially.

### 3.4. Other options to explore and future work

The innovation in the field of sintering during the last 10 years has been considerable [60], and several new sintering technologies have been brought into the picture, including cold sintering processes [117, 118], flash sintering [119, 120], and modified SPS methodologies [121, 122]. This widens the range of options for sintering ferrites with optimized microstructure and magnetic properties. Recently, Serrano *et al* have managed to produce dense SrM ceramic magnets with competitive magnetic properties using cold sintering routes that require much lower temperatures and reduce the energy consumption (>25%) [67, 68]. However, to the best of our knowledge, none of these other new sintering methods have been tested on hard hexagonal ferrites yet, which leaves plenty of room for an improvement in the research field analyzed in the present review.

## 4. Conclusions and perspectives

Hexagonal ferrites, in particular the M-type ($BaFe_{12}O_{19}$, $SrFe_{12}O_{19}$), are well-established PM materials, and they show great potential as gap magnets, in the run for substitution of rare-earths in some PM applications. The problems ecounterd in the consolidation of hexaferrites are outlined, and literature on the sintering of M-type ferrites is reviewed and discussed in the context of the density and magnetic properties of the resulting material.

The sintered magnets available in the market are prepared by traditional/conventional sintering methods. The conventional approaches require prolonged times and elevated temperatures, which promote an exaggerated growth of the ferrite grains which damages the coercivity of the densified piece. However, after years of investigation on the topic, addition of sintering aids has proven satisfactory in controlling this growth, yielding dense pieces with good magnetic properties.

Despite the good results of conventional sintering, great efforts have been and still are dedicated towards finding new or modified ways of densification of ferrite magnets, e.g. microwave sintering, SPS, flash sintering, cold sintering. This innovative methods reduce the sintering times and temperatures and/or the energy consumption by using faster and more efficient heating processes (MW, Joule heating) or experimental setups that allow for simultaneous application of pressure and temperature. Ferrite magnets sintered using these novel techniques are already yielding good densities and competitive magnetic properties, but the continuous development of the sintering technologies leaves space for further investigations in the field.

The latter is of the greatest importance since already small improvements in ferrites could lead to substitution of rare-earth-based magnets for a large number of applications, as in motors and generators even a mild remanence enhancement can lead to an increased power output, therefore paving the way to a cheaper, and more importantly, greener electrification. It is worth noting that the European Union wishes to reach climate neutrality by 2050 [123], and environmentally friendly technologies are vital for accomplishing this, giving a head start to magnets with less or no rare-earths. In addition, oxides as they are, ferrites are easily recyclable materials, providing a good starting point for a circular economy of such PMs.

## Data availability statement

No new data were created or analysed in this study.

## Acknowledgments

This work is supported by the European Commission through the H2020 project with Grant agreement H2020-NMBP-2016-720853 (AMPHIBIAN) and by the Spanish Ministerio de Ciencia, Innovación y Universidades (RTI2018-095303-A-C52). C G M acknowledges financial support from Spanish Ministerio de Ciencia e Innovación (MICINN) through the 'Juan de la Cierva' Program (FJC2018-035532-I). Financial support from the Slovenian Research Agency (research core funding No. P2-0087) is acknowledged.

## ORCID iDs

Cecilia Granados-Miralles  https://orcid.org/0000-0002-3679-387X

Petra Jenuš  https://orcid.org/0000-0002-1926-479X